\documentclass[rnote]{aa}  
\usepackage{graphicx}
\usepackage{color}
\usepackage{epsfig}
\usepackage{supertabular}
\usepackage{aalongtable}
\usepackage{txfonts}
\begin{document}
   \title{Compact stellar systems in the Fornax cluster: a UV perspective\thanks{Based on observations made with the NASA Galaxy
Evolution Explorer GALEX. GALEX is operated for NASA by the
Califonia Institute of Technology under NASA contract
NAS5-98034.}}

   %\subtitle{}

   \author{Steffen Mieske
          \inst{1}
          \and
          Michael Hilker\inst{1} \and Dominik J. Bomans\inst{2} \and Soo-Chang Rey\inst{3} \and Suk Kim\inst{3} \and { Suk-Jin Yoon}\inst{4} \and { Chul Chung}\inst{4}
          }

   \offprints{S. Mieske}

   \institute{European Southern Observatory, Karl-Schwarzschild-Strasse 2, 85748 Garching bei M\"unchen, Germany\\
              \email{smieske@eso.org}
         \and
     Astronomical Institute,
Ruhr-University Bochum,
Universit\"atsstr. 150,
44780 Bochum,
Germany
%\email{bomans@astro.ruhr-uni-bochum.de}
\and
     Department of Astronomy and Space Science, Chungnam National University, Daejeon 305-764, Korea
%\email{screy@cnu.ac.kr}
\and
Department of Astronomy \& Center for Space Astrophysics, Yonsei University, Korea
%\email{sjyoon@galaxy.yonsei.ac.kr}
}

   \date{}

% \abstract{}{}{}{}{} 
% 5 {} token are mandatory
 
  \abstract
  % context heading (optional)
   {In recent years, increasing evidence for chemical complexity and
multiple stellar populations in massive globular 
clusters (GCs) has emerged, including extreme horizontal branches (EHBs) and UV excess.} %leave it empty if necessary  
  % aims heading (mandatory)
{Our goal is to improve our understanding of UV excess in compact
  stellar systems, covering the regime of both
  ultra-compact dwarf galaxies (UCDs) and massive GCs.}
  % methods heading (mandatory)
{We use deep archival GALEX data of the central Fornax cluster to
  measure NUV and FUV magnitudes of UCDs and massive GCs. }
  % results heading (mandatory)
{We obtain NUV photometry for a sample of 35 compact objects that
  cover a range $-13.5<M_V<-10$ mag. Of those, 21 objects also have
  FUV photometry.  Roughly half of the sources fall into the UCD
  luminosity regime ($M_V\lesssim-$11 mag). We find that seven out of
  17 massive Fornax GCs exhibit a NUV excess with respect to
  expectations from stellar population models, both for models with
  canonical and with enhanced Helium abundance. This suggests that not
  only He-enrichment has contributed to forming the EHB population of
  these GCs.  The GCs extend to stronger UV excess than GCs in M31 and
  massive GCs in M87, at the 97\% confidence level. Most of the UCDs
  with FUV photometry also show evidence for UV excess, but their UV
  colours can be matched by isochrones with enhanced Helium abundances
  and old ages 12-14 Gyrs.
  We find that Fornax compact objects with X-ray emission detected
  from Chandra images are almost disjunct in colour from compact objects with
  GALEX UV detection, with only one X-ray source among the 35 compact
  objects.  However, since this source is one of the three most UV
  bright GCs, we cannot exclude that the physical processes causing
  X-ray emission also contribute to some of the observed UV
  excess.}
{}


   \maketitle 
%
%________________________________________________________________

\section{Introduction}
\label{introduction}
In recent years, increasing evidence for chemical complexity and
multiple stellar populations in several massive Galactic globular
clusters (GCs) has emerged (e.g. Gratton et al. 2004, Bedin et al.
2004, Piotto et al. 2007, Milone et al. 2008). For some globular
clusters (e.g. $\omega$\,Centauri and NGC 2808) there is evidence that
these multiple populations may be due to an enhancement in Helium
content (up to $Y\simeq 0.4$), as deduced from the presence of
extremely blue horizontal branches (EHBs) and multiple main sequences
(D'Antona et al. 2005, Lee et al. 2005).   The link between
  occurence of EHBs and He-enhancement is, however, under debate. EHB
  stars may also originate directly from fast rotating stars with an
enhanced mass loss during the red giant branch (RGB) evolution. Those
stars leave the RGB before the He flash towards the (He-core) white
dwarf cooling curve where they ignite Helium and end up on the EHB
(Castellani \& Castellani 1993, D'Cruz et al. 1996, Brown et al.
2001). This is sometimes called the 'hot-flasher' scenario. A
significant fraction of the EHB stars in $\omega$\,Centauri seem to
have followed this evolutionary path (Moehler et al. 2007).

It has been shown that EHB stars contribute most of the light in
the UV bands (e.g.  NGC 2808: Dieball et al. 2005). The presence of an
EHB in extragalactic - unresolved - GCs reveals itself in the
integrated light by a UV-excess compared to GCs with a `normal'
horizontal branch (HB). For example, Rey et al. (2007) find three
metal-rich ([Fe/H]$>-$1) GC candidates in M31 with significant FUV
flux which are thought to be analogs of two peculiar Galactic GCs, NGC
6388 and NGC 6441 (Yoon et al. 2008).  Sohn et al. (2006) and Kaviraj
et al. (2007) analysed the UV properties of massive globular clusters
associated with M87 in the Virgo cluster, and found that many of them
show a UV-excess with respect to canonical stellar population
models. These findings support the idea that EHBs may be a common
feature to the most massive compact stellar systems.

In this Research Note we focus on the UV properties of compact
stellar systems in the Fornax cluster. In contrast to the studies of
Sohn et al. and Kaviraj et al. on Virgo GCs, we extend our analysis to
compact stellar systems beyond the mass range of GCs ($M\lesssim 3
\times 10^6$ M$_{\sun}$), including the so-called ultra-compact dwarf
galaxies (UCDs, Drinkwater et al. 2003), which cover the mass range up
to $\sim10^8$M$_{\sun}$, having $M_V \lesssim -11$ mag.
We analyse how the UV properties of UCDs compare to
those of both massive and normal GCs, in order to improve our
  knowledge of EHB occurence in compact stellar systems.  Throughout
this study we adopt (m-M)=31.4 mag (Freedman et al. 2001) as distance
modulus to the Fornax cluster.
\begin{figure*}[]
\begin{center}
  \epsfig{figure=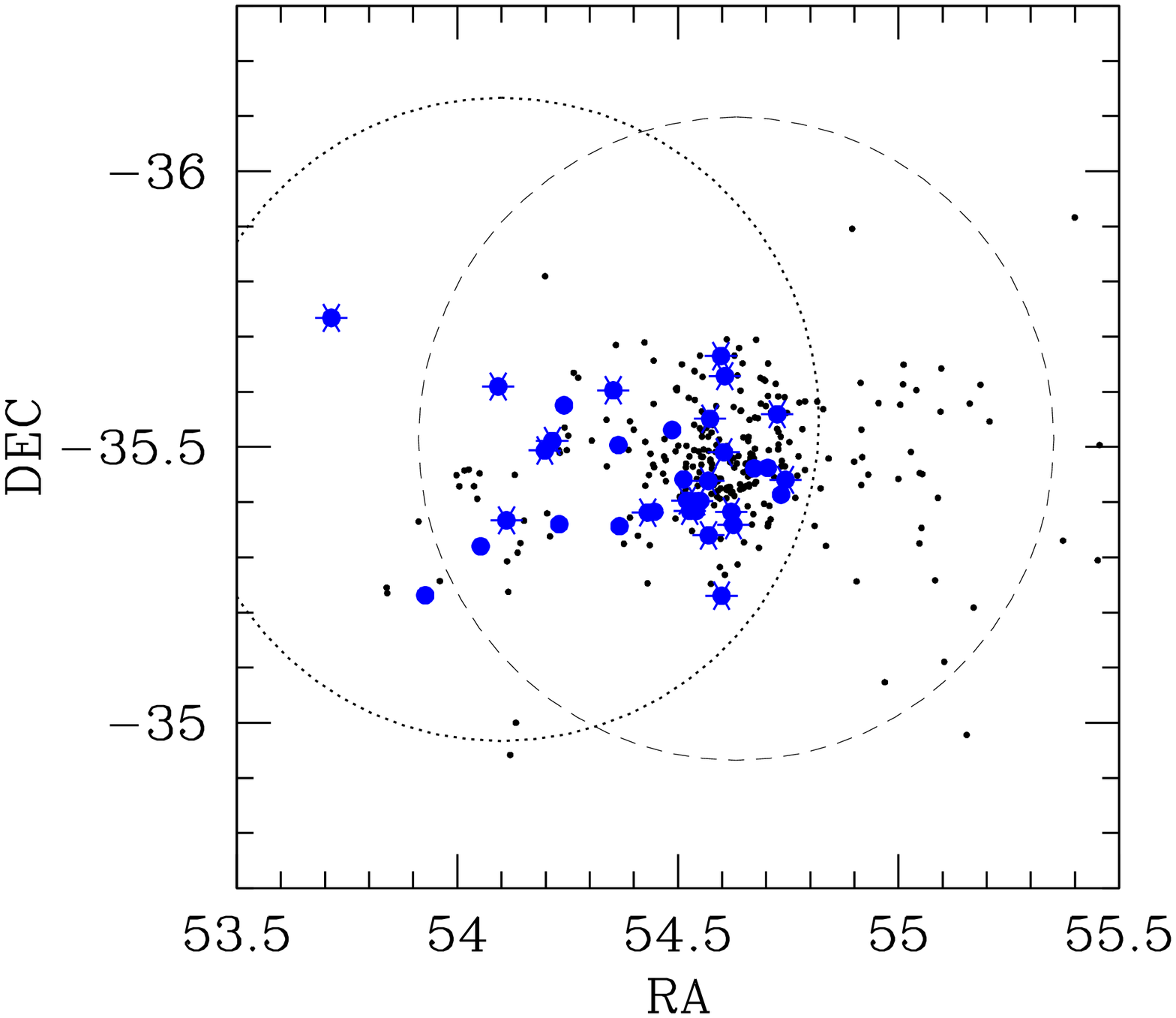,width=8.6cm}
  \epsfig{figure=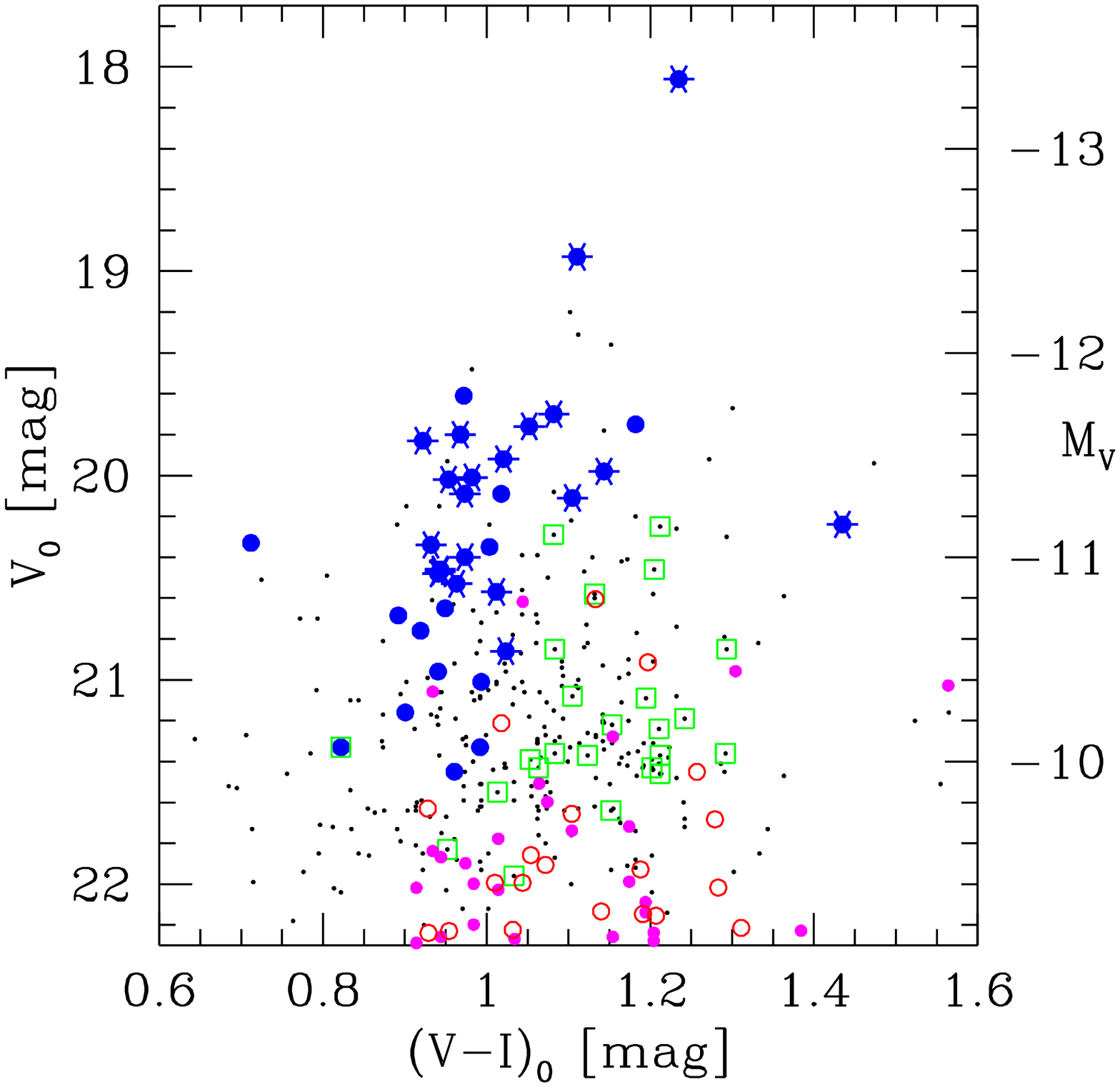,width=8.6cm}

  \caption{{\bf Left:} Map of the central Fornax cluster. The large circles indicate the FoV of the two archival GALEX pointings used for this study. The large dotted circle corresponds to the Deep Imaging Survey (DIS), the large dashed circle corresponds to the Near Galaxies Survey (NGS). The small dots indicate our
sample of spectroscopically confirmed compact objects down to $V\simeq 22$ mag ($M_V=-9.4$ mag). The filled circles indicate the sources with visually verified GALEX matches in the NUV, the asterisks are those sources with matches in the FUV (matching radius 3$''$). {\bf Right:} V,I colour-magnitude diagram of the same sources as in the left panel. The optical data is from Mieske et al.  (2004 \& 2006, 2007), and Dirsch et
al. (2003). Magenta dots indicate GCs in M31 with UV colours available (Fig.~\ref{cmds}; Rey et al. 2007), red circles indicate GCs in M87 with FUV colours available (Fig.~\ref{cmds}; Sohn et al. 2006). Green squares indicate sources with
X-ray counterparts from the Chandra study of Scharf et al. (2005). Note that only one of the GALEX UV detections has a detected X-ray counterpart.}

\label{map}
\end{center}
\end{figure*}

\section{The data}
\label{data}
In order to investigate the UV properties of massive compact stellar
systems in Fornax, we retrieved archival FUV and NUV GALEX images from
two pointings in the Fornax
cluster\footnote{http://galex.stsci.edu/GR2/}.  One pointing was
offset from NGC 1399 by about 0.4$\degr$ to the west, and had been
taken within the Deep Imaging Survey (DIS) with integration time 18000
seconds. The other pointing was centered on NGC 1399, taken within the
Near Galaxy Survey (NGS), with integration time 1700 seconds in FUV
and NUV.  See Fig.~\ref{map}.

We subtracted the two giant elliptical galaxies NGC 1399 and NGC 1404
from the archival GALEX images, using the modelling routines
\texttt{ellipse} and \texttt{bmodel} within IRAF. Then we run
SExtractor on the images to create a source catalog of detections in
NUV and FUV. For this we required a minimum of 5 adjacent pixels with
at least 2$\sigma$ above the sky noise. We adopted MAG\_BEST as source
magnitude, and used the GALEX photometric zero-points given in the
image headers.  From artificial star experiments using the same
  detection parameters we derived 50\% completeness limits for
  unresolved sources in the DIS images of NUV$_0=24.2$ mag, and
  FUV$_0=24.7$ mag. The region within $\simeq$2$'$ of the center of
  NGC 1399 showed a considerably brighter completeness limit by 1-2
  mag. The GALEX detections in the output catalogs were then matched
with the position of compact Fornax cluster members known from an
up-to-date compilation of literature spectroscopic surveys in Fornax,
which extends to about V$\simeq$22.2 mag (Kissler-Patig et al.
  1999; Mieske et al. 2002 \& 2004; Dirsch et al. 2004; Bergond et al.
  2007; Firth et al. 2007; Richtler et al. 2008; Karick et al. 2008,
  private communications; see also Tables~\ref{table1}
  and~\ref{table2}). There are no compact Fornax cluster members
  known within $\simeq$2$'$ of NGC 1399 (Fig.~\ref{map}), such that
  the completeness drop in this region is irrelevant in the context of
  this study. The allowed matching radius was 3$''$, which is about 2
GALEX pixels, or 2/3 of the GALEX PSF FHWM.

All matches on the GALEX images were visually classified in an
independent manner by the authors SM and SCR into clear and marginal
detections.  We kept matches if they were classified as clear
detection by both authors. In addition we also kept matches in a given
filter band for which at least one of the authors gave a clear
classification, if the source was classified by both authors as a
clear detection in the other filter band.  We excluded sources from
the match list which from higher resolution optical imaging (Mieske et
al. 2007) had neighbouring sources within a radius of 4$''$ that were
not fainter than $V=V_{\rm obj}+3$ mag. This was to make sure that the
detected UV flux indeed originates from the compact object and not a
close neighbour.  About 20\% of the UV matches were affected by this
rejection.

The final sample of visually confirmed NUV matches contains 35 objects
with $18<V<21.4$ mag ($-13.4<M_V<-10$ mag), while the catalog of FUV
matches contains 21 objects with $18<V<20.8$ mag ($-13.4<M_V<-10.6$
mag). All of the FUV matches are also NUV matches. In
  Tables~\ref{table1} and ~\ref{table2} the photometric properties of
  the objects are listed. Fig.~\ref{map} shows a map of the
investigated area and a V,I colour-magnitude diagram, indicating the
FUV and NUV matches together with the full literature sample of Fornax compact
objects. The photometry was corrected for foreground extinction using
the reddening maps of Schlegel et al.  (1998). Mainly optically
  blue GCs are detected in the GALEX images. We attribute this to the
  fact that the completeness limits of this data set favour the
  detection of UV bright sources in the GC magnitude regime (see next
  Section). In Fig.1 (right panel) we also indicate data points for GCs in M31 and M87 with
UV coverage.

\begin{figure*}[]
\begin{center}
  \epsfig{figure=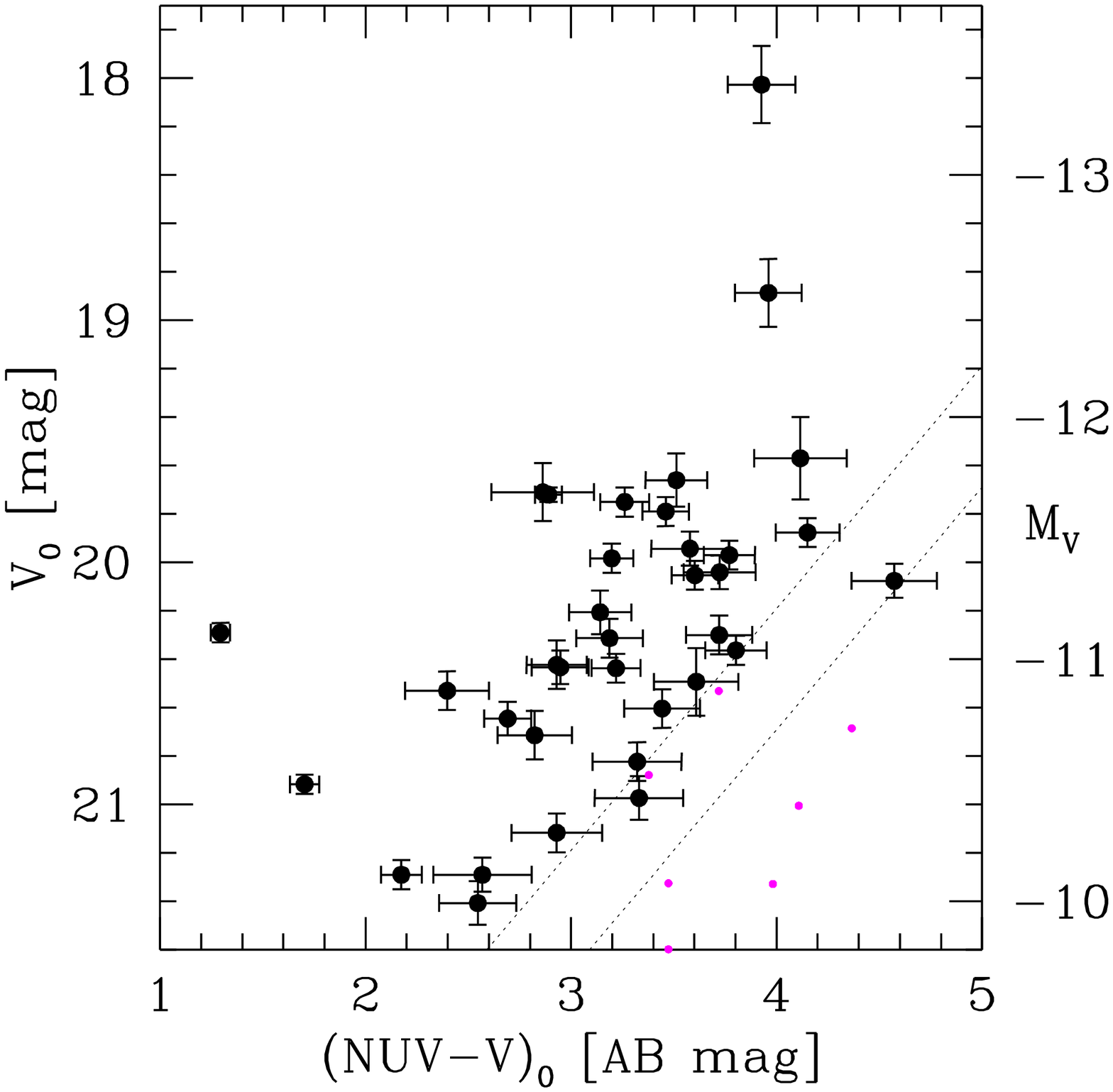,width=8.6cm}
  \epsfig{figure=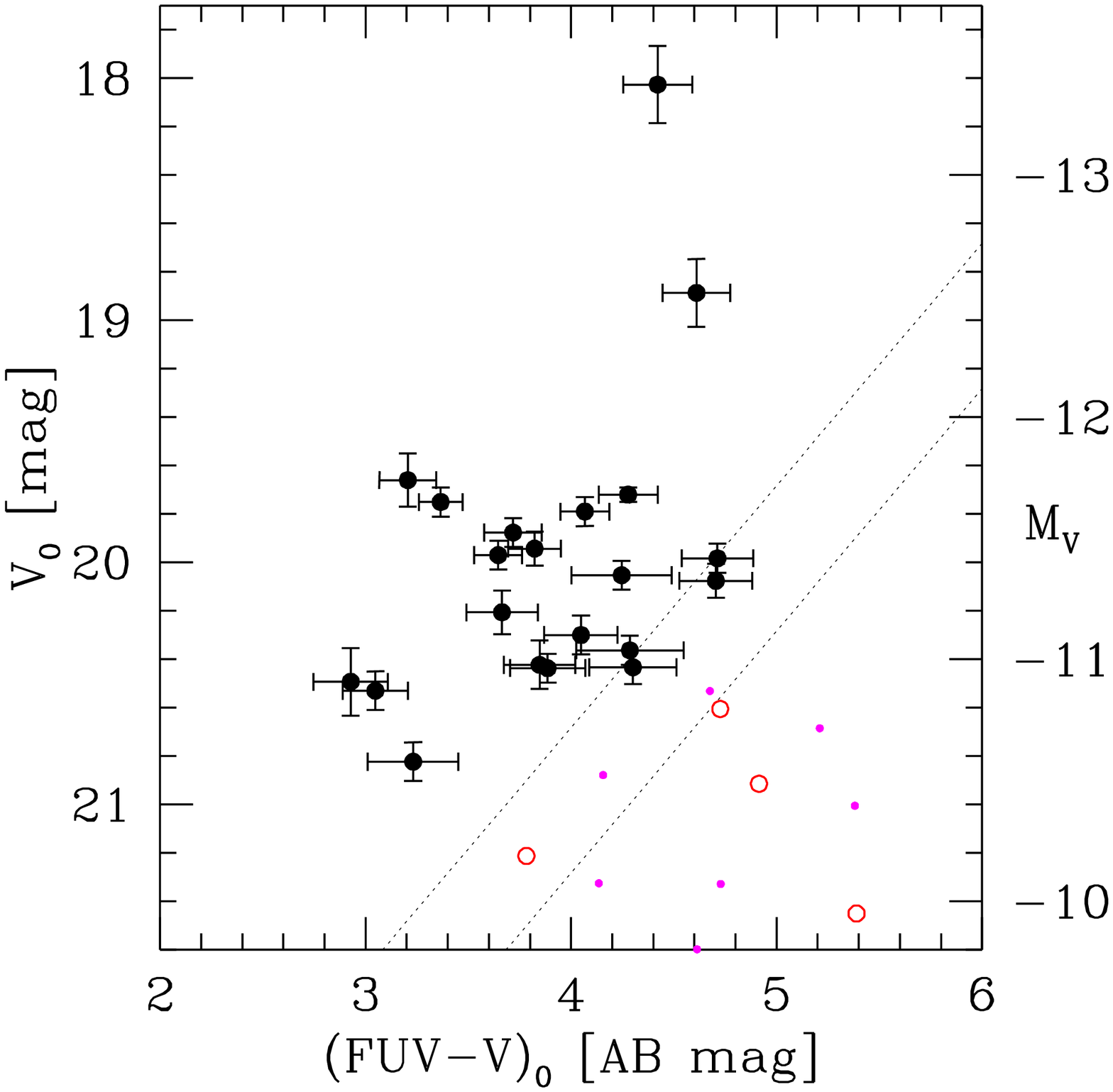,width=8.6cm}
  \caption{{\bf Left panel:} NUV-V colour-magnitude diagram of compact objects
    in the Fornax cluster with matches in the archival GALEX images
    (see text). Note the possible break at about $M_V\simeq -11.1$
    mag. Brighter sources are exclusively red, with (NUV-V)$>$2.7 mag.
    The two diagonal dotted lines indicate the 50\% and 90\%
    completeness limit of the GALEX detections{ , as derived from artificial star experiments.}  Small magenta dots indicate data
    points for GCs in M31 (Rey et al.~2007). {\bf Right panel:} FUV-V
    CMD of compact objects in the Fornax cluster with with matches in
    the archival GALEX images (see text). Small magenta dots indicate data points for
    GCs in M31 (Rey et al.~2007). Red circles indicate data points for
    GCs in M87 (Sohn et al. 2006).}

\label{cmds}
\end{center}
\end{figure*}

\begin{figure*}[]
\begin{center}
  
  \epsfig{figure=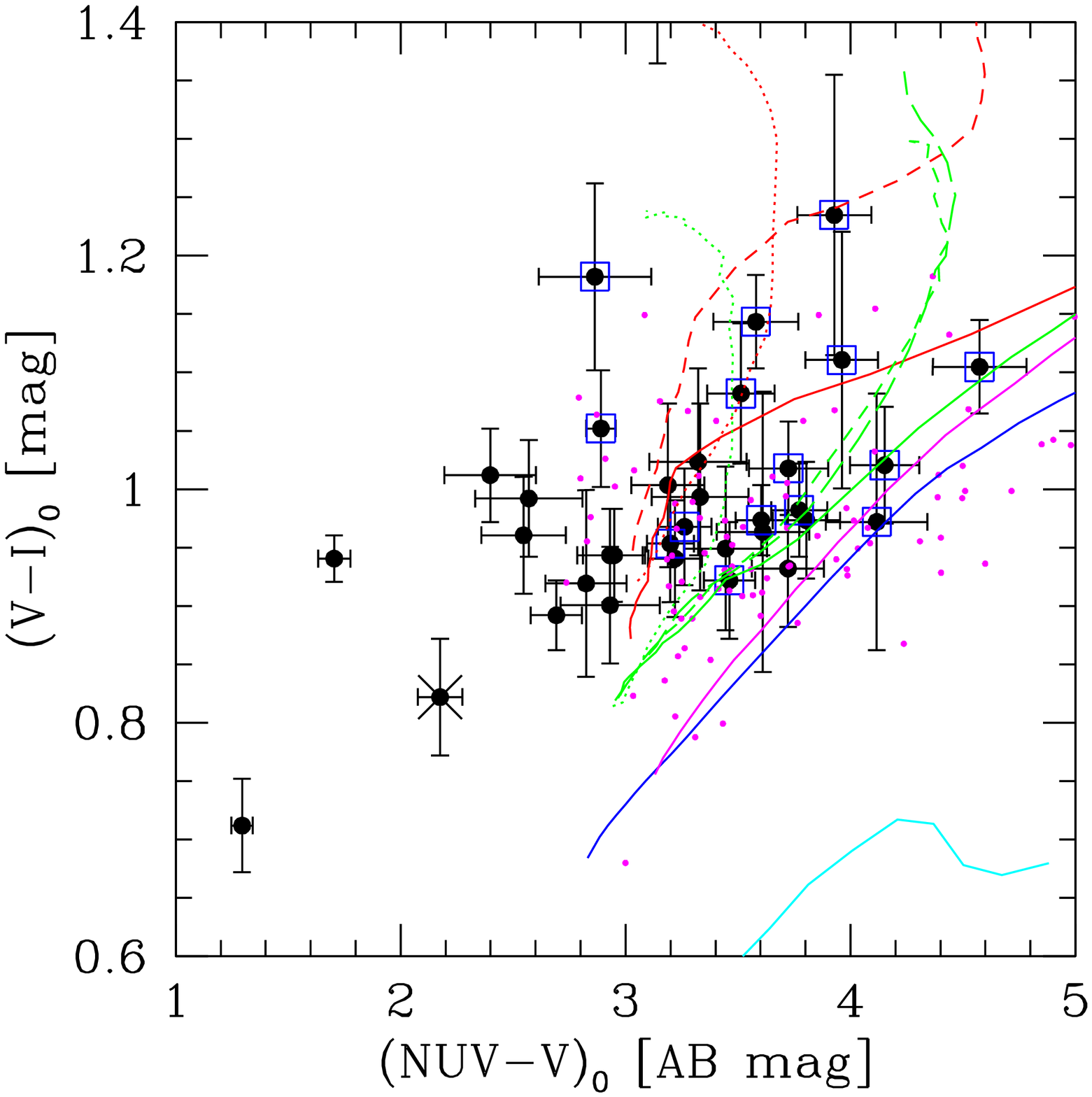,width=8.5cm}
  \epsfig{figure=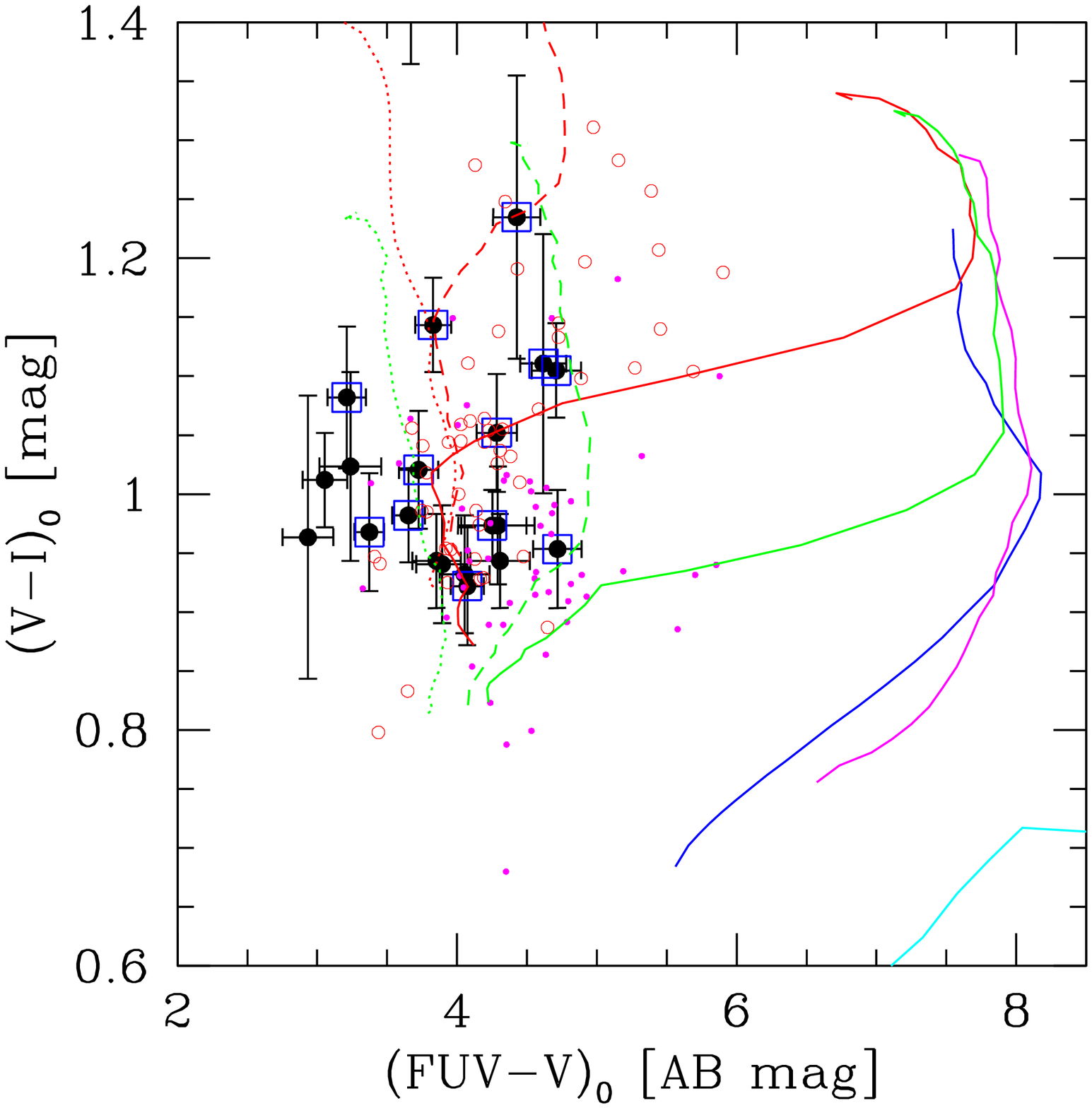,width=8.5cm}
  \epsfig{figure=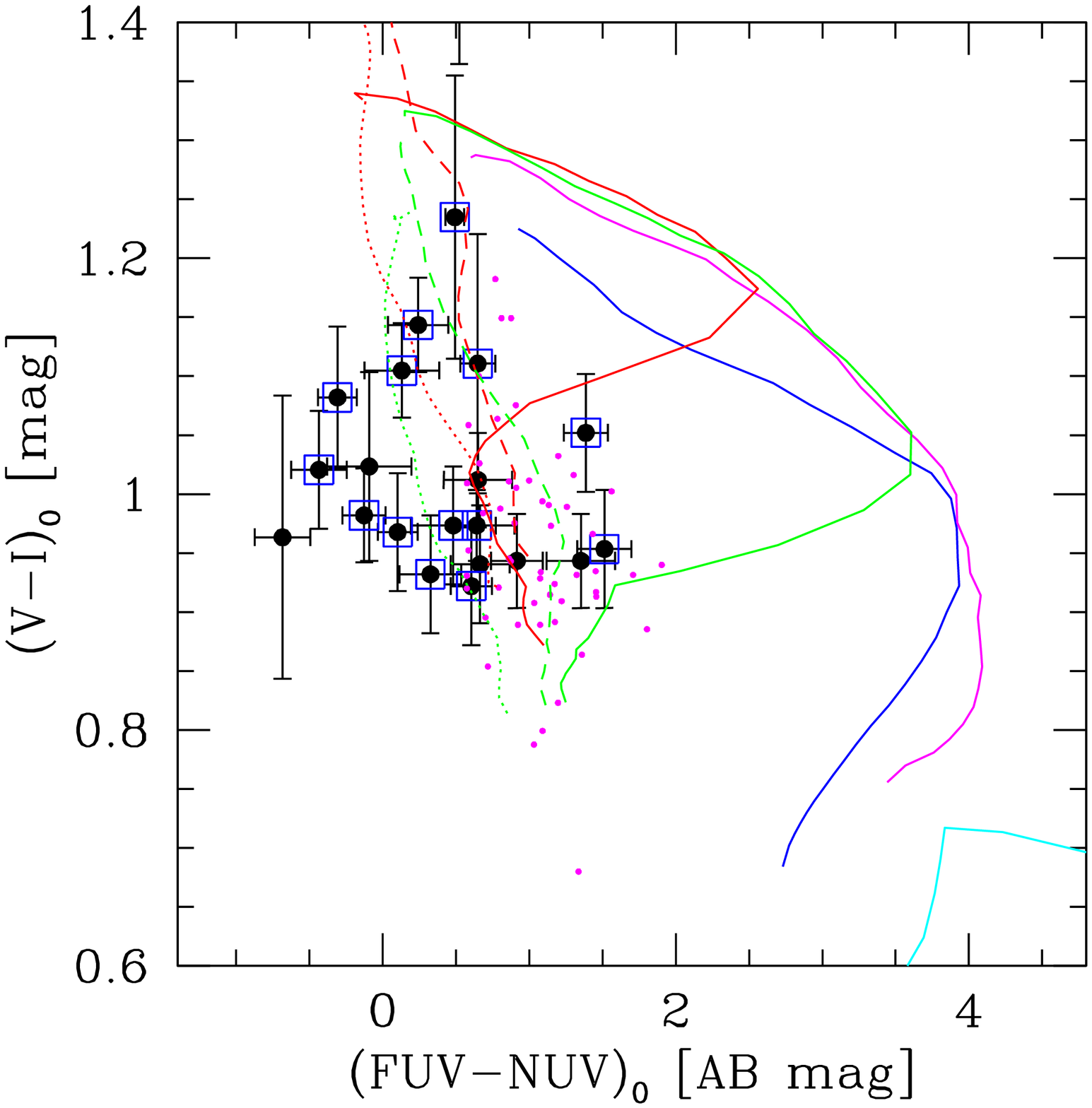,width=8.5cm}
  \epsfig{figure=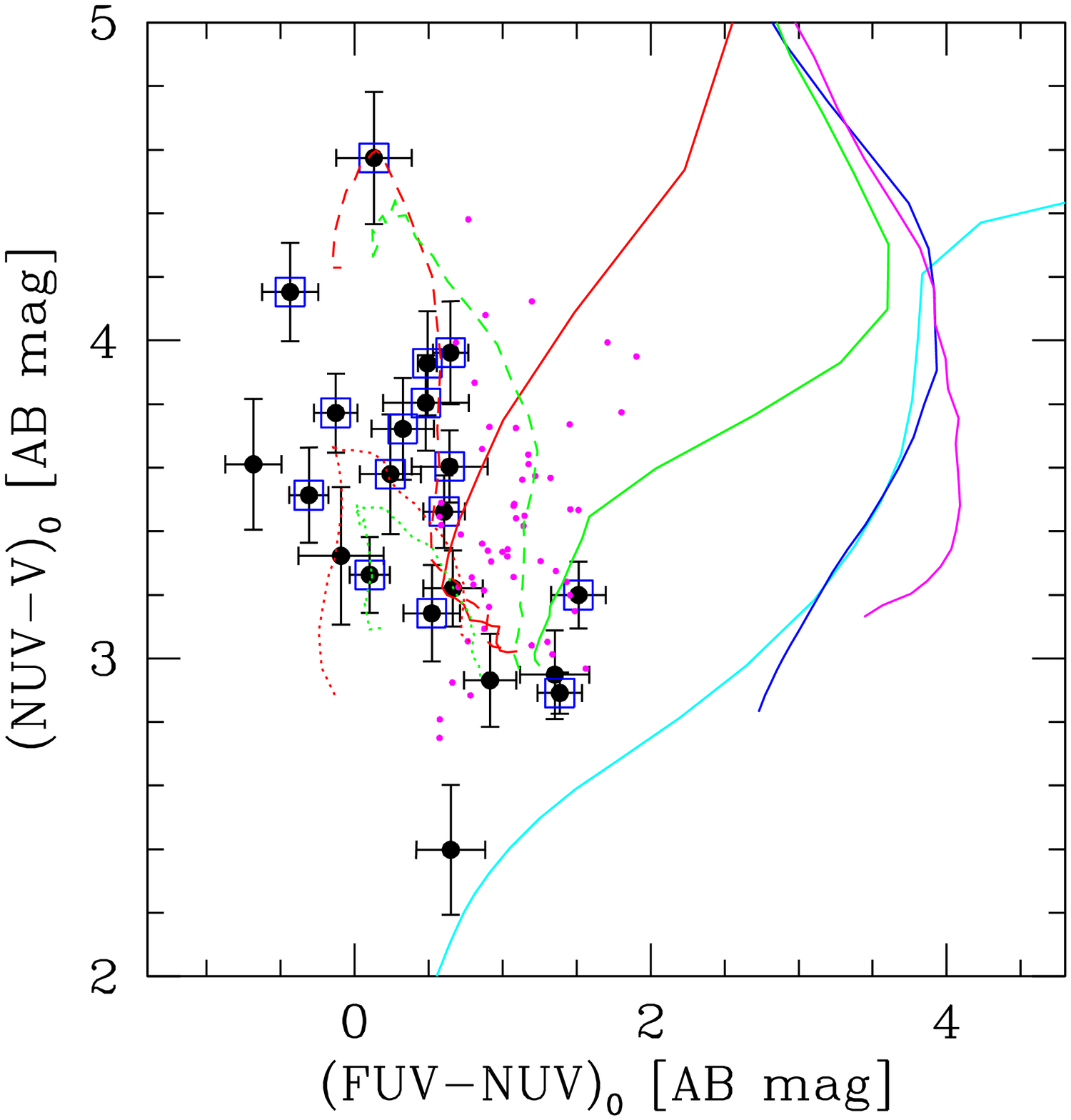,width=8.5cm}

  \caption{{\bf Top left panel:} Colour-colour diagram of (NUV-V) vs. (V-I) for the objects from Fig.~\ref{cmds}. UCDs, i.e. objects with $M_V<-11.1$ mag, are marked by (blue) squares. Overlaid in solid lines are model isochrones from the { YEPS models (see text for details) }for metallicities between [Fe/H]=$-$2.5 and 0.5 dex{ , and assuming [$\alpha$/Fe]=0.3 dex}. The isochrones correspond to the following ages: 1 Gyr (cyan), 5 Gyr (blue), 9 Gyrs (magenta), 12 Gyrs (green), 14 Gyrs (red).  All solid lines assume the standard Helium
    abundance Y=0.23. The dashed (dotted) lines indicate the 12
    and 14 Gyr isochrones for a fraction of 30\% (100\%) of
    stars with Helium abundance Y=0.34. The (green) long dashed
      line shows the 12 Gyr isochrone with a 30\% contribution of
      stars with Y=0.40. There is almost no difference between Y=0.34
      and Y=0.40. The source marked by a cross is the only GALEX UV
    detection that has an X-ray counterpart in the Chandra X-ray
    images (Scharf et al. 2005). {\bf Top right panel:} Colour-colour
    diagram of (FUV-V) vs. (V-I). Model tracks as in the left panel.
    Red circles indicate data points for GCs in M87 (Sohn et al.
    2006). {\bf Bottom left panel:} Colour-colour diagram of (FUV-NUV)
    vs.  (V-I). Model tracks as in the upper two panels. {\bf Bottom
      right panel:} Colour-colour diagram of (FUV-NUV) vs. (NUV-V).
    Model tracks as in the upper two panels. }
\label{colcol}
\end{center}
\end{figure*}

\begin{table*}
\caption{{ Photometric properties of the 21 compact Fornax cluster members detected in both NUV and FUV in the GALEX archival images. The targets are ordered by increasing V$_0$ magnitude. The object IDs indicate the following sources: UCDxx = Firth et al. (2007) \& Evstigneeva et el. (2008); FCOS xxxxx = Mieske et al. (2002, 2004); Yxxxxx = Richtler et al. (2008); gcxxx and ucxxx = Bergond et al. (2007);  Kxxxx = Karick et al. (in preparation, private communications).}}
\begin{center}
\begin{tabular}{|l|rrrrrrr|}
\hline
ID & RA [2000] & DEC [2000]&V$_0$ [mag] & (V-I)$_0$  &(NUV-V)$_0$ &(FUV-V)$_0$ &(FUV-NUV)$_0$ \\\hline
UCD3 &  3:38:54.0 & -35:33:33.8 &  18.03 (0.16) &  1.23 (0.12) &  3.93 (0.16) &  4.42 (0.17)  & 0.49 (0.06) \\
UCD6 &  3:38:05.0 & -35:24:09.7 &  18.89 (0.14) &  1.11 (0.11) &  3.96 (0.16) &  4.61 (0.17)  & 0.65 (0.12) \\
UCD27 &  3:38:10.3 & -35:24:06.1 &  19.66 (0.11) &  1.08 (0.06) &  3.51 (0.15) &  3.21 (0.14)  & $-$0.31 (0.13) \\
UCD12 &  3:36:26.7 & -35:22:01.6 &  19.72 (0.03) &  1.05 (0.05) &  2.89 (0.07) &  4.28 (0.14)  &  1.39 (0.15) \\
UCD25 &  3:37:43.6 & -35:22:52.0 &  19.75 (0.06) & 0.97 (0.05) &  3.26 (0.12) &  3.37 (0.11)  & 0.10 (0.14) \\
UCDk &  3:38:23.7 & -35:13:49.4 &  19.79 (0.06) & 0.92 (0.05) &  3.46 (0.11) &  4.07 (0.12)  & 0.60 (0.14) \\
K1011 &  3:37:24.8 & -35:36:10.1 &  19.88 (0.06) &  1.02 (0.05) &  4.15 (0.15) &  3.72 (0.14)  & $-$0.43 (0.19) \\
UCD41 &  3:38:29.0 & -35:22:56.6 &  19.94 (0.07) &  1.14 (0.04) &  3.58 (0.19) &  3.82 (0.13)  & 0.24 (0.21) \\
UCDm &  3:38:06.5 & -35:23:03.8 &  19.97 (0.06) & 0.98 (0.04) &  3.77 (0.12) &  3.65 (0.12)  & $-$0.13 (0.15) \\
K1002 &  3:36:22.2 & -35:36:34.6 &  19.98 (0.06) & 0.95 (0.05) &   3.20 (0.10) &  4.71 (0.17)  &  1.51 (0.18) \\
UCD32 &  3:38:16.7 & -35:20:23.3 &  20.05 (0.06) & 0.97 (0.03) &   3.60 (0.11) &  4.25 (0.24)  & 0.64 (0.26) \\
UCD39 &  3:38:25.5 & -35:37:42.6 &  20.08 (0.07) &   1.10 (0.04) &  4.57 (0.21) &   4.70 (0.18)  & 0.13 (0.25) \\
uc218.7 &  3:38:23.4 & -35:39:53.3 &  20.21 (0.09) &  1.43 (0.07) &  3.14 (0.15) &  3.66 (0.17)  & 0.52 (0.19) \\
UCD31 &  3:38:16.5 & -35:26:19.3 &   20.30 (0.08) & 0.93 (0.05) &  3.72 (0.16) &  4.05 (0.18)  & 0.33 (0.21) \\
UCD33 &  3:38:17.5 & -35:33:04.0 &  20.36 (0.06) & 0.97 (0.05) &   3.80 (0.15) &  4.29 (0.26)  & 0.48 (0.29) \\
Y99025 &  3:38:58.6 & -35:26:26.2 &  20.42 (0.10) & 0.94 (0.04) &  2.93 (0.15) &  3.85 (0.17)  & 0.92 (0.18) \\
K1007 &  3:36:47.6 & -35:29:37.3 &  20.43 (0.07) & 0.94 (0.04) &  2.95 (0.14) &   4.30 (0.21)  &  1.35 (0.23) \\
K1000 &  3:34:51.4 & -35:44:02.8 &  20.44 (0.06) & 0.94 (0.05) &  3.22 (0.12) &  3.89 (0.18)  & 0.67 (0.20) \\
UCD38 &  3:38:25.1 & -35:29:25.1 &  20.49 (0.14) & 0.96 (0.12) &  3.61 (0.21) &  2.93 (0.18)  & $-$0.68 (0.19) \\
UCD17 &  3:36:51.7 & -35:30:38.9 &  20.53 (0.08) &  1.01 (0.04) &   2.40 (0.20) &  3.05 (0.16)  &  0.65 (0.23) \\
FCOS 0\_2032 &  3:38:30.2 & -35:21:31.0 &  20.82 (0.08) &  1.02 (0.08) &  3.32 (0.22) &  3.23 (0.22)  & $-$0.09 (0.29) \\\hline

\end{tabular}
\label{table1}
\end{center}
\end{table*}

\begin{table*}
\caption{{ Photometric properties of the 14 compact Fornax cluster members detected only in the NUV GALEX archival images. The targets are ordered by increasing V$_0$ magnitude. The object IDs indicate the following sources: NTTxxx = Kissler-Patig et al. (1999); UCDxx = Firth et al. (2007) \& Evstigneeva et el. (2008); FCOS xxxxx = Mieske et al. (2002, 2004); Yxxxxx = Richtler et al. (2008); gcxxx and ucxxx = Bergond et al. (2007);  xx.xxx = Dirsch et al. (2004).}}
\begin{center}
\begin{tabular}{|l|rrrrr|}
\hline
ID & RA [2000] & DEC [2000]&V$_0$ [mag] & (V-I)$_0$  &(NUV-V)$_0$ \\\hline
NTT414 &  3:38:09.7 & -35:23:01.3 &  19.57 (0.17) & 0.97 (0.11) &  4.12 (0.23) \\
NTT410 &  3:38:12.2 & -35:24:06.5 &  19.71 (0.12) &  1.18 (0.08) &  2.86 (0.25) \\
UCD20 &  3:37:27.6 & -35:30:12.6 &  20.04 (0.07) &  1.02 (0.04) &  3.72 (0.18) \\
UCD18 &  3:36:55.5 & -35:21:36.0 &  20.29 (0.04) & 0.71 (0.04) &  1.29 (0.05) \\
FCOS 1\_0630 &  3:38:56.1 & -35:24:49.0 &  20.31 (0.08) &     1.00 (0.07) &  3.19 (0.16) \\
FCOS 2\_2165 &  3:37:28.2 & -35:21:23.0 &   20.60 (0.08) & 0.95 (0.07) &  3.44 (0.19) \\
gc235.7 &  3:36:12.7 & -35:19:11.6 &  20.64 (0.07) & 0.89 (0.03) &  2.69 (0.11) \\
Y7225 &  3:37:47.2 & -35:22:57.4 &  20.71 (0.10) & 0.92 (0.08) &  2.82 (0.18) \\
gc290.6 &  3:35:42.5 & -35:13:52.0 &  20.92 (0.04) & 0.94 (0.02) &   1.70 (0.07) \\
FCOS 1\_2089 &  3:38:48.9 & -35:27:43.9 &  20.97 (0.09) & 0.99 (0.08) &  3.33 (0.22) \\
91.041 &  3:37:56.9 & -35:31:50.9 &  21.12 (0.08) & 0.90 (0.05) &  2.93 (0.22) \\
86.112 &  3:38:41.4 & -35:27:41.0 &  21.29 (0.07) & 0.99 (0.05) &  2.57 (0.24) \\
Y1149 &  3:36:58.0 & -35:34:32.2 &  21.29 (0.06) & 0.82 (0.05) &  2.17 (0.10) \\
81.008 &  3:38:03.0 & -35:26:28.3 &  21.41 (0.09) & 0.96 (0.05) &  2.55 (0.19) \\\hline

\end{tabular}
\label{table2}
\end{center}
\end{table*}

\section{Results}

In Fig.~\ref{cmds} we plot two colour-magnitude diagrams (CMDs) of the
NUV matches, one of (NUV-V) vs. V, and one of (FUV-V) vs. V. { As
  stated before, all the FUV matches are also NUV matches.} We
indicate the magnitude dependent colour limits of the data, which
biases us towards detecting UV bright objects at fainter optical
luminosities ({see also Sect.~\ref{data})}. In Fig.~\ref{colcol},
  we plot colour-diagrams of (V-I) vs. (NUV-V), (V-I) vs. (FUV-V),
  (V-I) vs.  (FUV-NUV), and (NUV-V) vs.  (FUV-NUV). In all four plots
  we also indicate the GALEX data for GCs in M31, taken
  from the compilation of Rey et al.~(2007). Furthermore, we show FUV
  data points for the M87 GCs from the compilation of Sohn et al.
  (2006). Note that due to the sensitivity limit of our used GALEX
  data, we would at the distance of Fornax be able to detect only two
  of the M31 GCs, and none of the M87 GCs (see Fig.~\ref{cmds}).

  In the colour-magnitude diagram of (NUV-V) vs. V { (Fig.~\ref{cmds},
    left panel)}, we find that sources brighter than $M_V\simeq$$-$11.1
  mag exhibit exclusively ``red'' colours $(NUV-V)\gtrsim 2.7$ mag.
  Fainter than this, several sources extend to bluer colours
  $(NUV-V)\simeq 1.0$, indicating a UV excess relative to the brighter
  objects. Note that $M_V\simeq-$11 mag corresponds to the approximate
  separation between ordinary GCs and UCDs (Ha\c{s}egan et
  al. 2005; Mieske et al. 2006). We can therefore state that seven out
  of 17 massive Fornax GCs detected in the GALEX images exhibit a UV
  excess relative to UCDs.  Massive Fornax GCs ($M_V>-11.1$ mag) have
  a mean colour of (NUV-V)={ 2.92 $\pm$ 0.13 mag}, Fornax UCDs
  ($M_V<-11.1$ mag) have (NUV-V)=3.49 $\pm$ 0.17 mag, while the full
  sample of M31 GCs has a mean colour of (NUV-V)=3.73 $\pm$ 0.07 mag
  (see Fig.~\ref{colcol}).

Is this relative UV-excess indicative of more extreme horizontal
branches in those Fornax GCs? Or is the UV-excess simply due to a much
younger age?  { To address this question, we make use of a grid of
  simple stellar population (SSP) models which are constructed using
  the Yonsei Evolutionary Population Synthesis (YEPS) code (Park \&
  Lee 1997; Lee, Yoon \& Lee 2000; Lee et al. 2005; Yoon et al. 2006,
  2008). We note that the models in the present study are the latest
  version of the YEPS model. The version has adopted a new set of HB
  evolutionary tracks that were built using the identical input
  physics and equations of state as the Yonsei-Yale ($Y^{2}$) MS-RGB
  evolutionary tracks (Kim et al. 2002) and taking into account the
  $\alpha$-element enhancement effect. In Rey et al. (2005, 2007) and
  Kaviraj et al. (2007), the YEPS models were used to investigate the
  integrated light of GCs in M31 based on the GALEX UV
  data, and in the Virgo cluster (M87) based on HST/STIS UV data,
  respectively. In Lee et al. (2005) the properties of resolved
  stellar populations in $\omega$Cen and NGC 2808 were analysed using
  the YEPS models (see also Rey et al. 2001, 2004; and Yoon \& Lee 2002 for
  an application to resolved stellar populations in ``normal''
  GCs).}

Fig.~\ref{colcol} shows that massive Fornax GCs with $(NUV-V)<2.7$ mag
indeed exhibit a UV-excess with respect to the { SSP model}
predictions. There is no combination of age and metallicity which can
reproduce these very blue NUV colours at the given (V-I). Not even
isochrones with an enhanced He abundance (Y=0.34 instead of the
canonical value Y=0.23) can account for them. The UCD data points are
more consistent with the model tracks for old ages around 12-14 Gyrs.
There is a slight UV excess for the optically red UCDs, which can be
explained by He enhanced isochrones of old ages.

We now assess whether detecting some massive Fornax GCs with UV excess
indicates that Fornax GCs as a sample have a higher {\it probability}
for a UV excess than our comparison sample, the M31 GCs.  The two
samples are almost disjunct in luminosity (see Fig.~\ref{cmds}). We
focus on (NUV-V)$<$2.7 mag, which is the blue limit of the sample of
87 M31 GCs with NUV detection (Fig.~\ref{colcol}), and also
corresponds to the blue limit of the UCD colours. Out of a total of
173 massive GCs with $-11.1<M_V<-9.9$ mag in the GALEX FoV, only 7
(4$\pm$1.5\%) have (NUV-V)$<$2.7 mag. A Poisson test shows that
drawing 0 out of 87 at an underlying assumed probability of
$0.04^{+0.015}_{-0.015}$ occurs in 3$^{+8}_{-2.1\%}$\% of random
samplings. We can therefore state that massive Fornax GCs extend to
higher NUV fluxes than M31 GCs at the 97\% confidence level. Note that
also none of the 29 UCDs in the GALEX FoV has (NUV-V)$<$2.7 mag.
However, this non-detection is not statistically significant when
assuming an underlying probability of $0.04^{+0.015}_{-0.015}$ for a
NUV excess as deduced from the massive GCs.

In the (FUV-V) vs. (V-I) colour-colour diagram of Fig.~\ref{colcol}, a
much better age resolution is achieved, but the number of GALEX
detections drops to 21, as does the number of M31 GCs (49 instead of
87 for NUV).  In this diagram, a UV excess for the 2-3 most metal-rich
Fornax UCDs is confirmed, similar to the metal-rich M87 GCs.  Those
UCDs show colours best matched with He-enhanced isochrones of old ages
12-14 Gyrs. Due to the brighter
sensitivity limit of the FUV data, only seven massive GCs enter the
sample, of which three have UV excess marginally incompatible with
He-enhanced isochrones. Given the bias towards detecting UV bright
sources, and the brighter sensitivity limit in the FUV than in the
NUV, we do not find statistically significant evidence for a different (FUV-V)
distribution between Fornax UCDs and massive GCs or M31 GCs.

The (FUV-NUV) colour-colour diagrams in Fig.~\ref{colcol}, especially
the diagram (FUV-NUV) vs. (NUV-V), allow the best discrimination into
objects consistent and inconsistent with standard He abundance
isochrones. Here, Fornax UCDs with $(FUV-NUV)$$\lesssim$0.5 mag show
UV colours consistent with enhanced He abundance that cannot be
explained by standard He isochrones of 14 Gyrs.  They also show that
at a given colour (V-I) or (NUV-V), UCDs are bluer in (FUV-NUV) than
our comparison sample of M31 GCs. This is the strongest evidence in
our data for UCDs as a class harbouring stellar populations with UV
excess. Again, we cannot judge on a difference between UCDs and
massive Fornax GCs due to the brighter sensitivity limit in the FUV,
which excludes most of the GCs detected in the NUV.

\section{Discussion and conclusion}

A UV excess in an old stellar population is likely due to EHB stars.
As pointed out in Sect.~\ref{introduction}, an EHB may be linked
  to helium-enriched stars (e.g. Ventura et al. 2001, D'Antona et al.
  2002). The strong UV excess of the seven massive Fornax GCs beyond
  the He-enhanced isochrones, especially the NUV excess of the three most
  extreme GCs with (NUV-V)$<$2.4 mag (see Fig.~\ref{colcol}), suggests
  that in these objects, EHB formation is also driven by other
  processes. In this context, a plausible explanation may be enhanced
  mass loss of evolved stars, triggered by high stellar densities
  (Decressin et al.  2007; Huang \& Gies 2006) and/or large binary
  fractions.

Excess radiation at short wavelengths can in principle also arise from
accretion onto a black hole (King et al. 1993), which can be traced by
low-mass X-ray binaries (Jord\'{a}n et al. 2004).  We have
cross-checked the positions of all GALEX UV detections with X-ray
source detections in the Chandra Fornax Survey data (Scharf et al.
2005 and private communication), the deepest available wide-field
X-ray survey of Fornax (50ks integration with ACIS). The sensitivity
of these images is a few $ 10^{38}$ erg/sec, allowing to detect the
most luminous LMXBs (Jord\'{a}n et al.  2004).  In Fig.~\ref{map} {
  (right panel)} we indicate the (V-I) optical colours of those
compact objects with X-ray matches. 
At a given magnitude, the X-ray matches { in GCs} are biased
towards red optical colours (see also Jord\'{a}n et al.  2004), while
GALEX UV detections are biased towards 
blue optical colours.  This suggests that generally, the UV- and X-ray-emission
  of the compact stellar systems are not caused by the same physical
  processes.  However, there is one GALEX UV detection with an X-ray
  counterpart (Fig.~\ref{map} and~\ref{colcol}), which happens to be
  one of the three GCs with largest UV excess. We can therefore not
  exclude that the UV excess in some of the GCs is linked to
  accretion processes.

We finally note that comparing the probability of UV excess
  between UCDs and GCs allows to test whether EHBs are more likely
  associated with present-day deep potential wells (i.e. UCDs) or high
  stellar densities (i.e. GCs; Dabringhausen et al. 2008, Mieske et
  al. 2008). One would expect deep potential wells to favour
  self-enrichment (e.g. Ventura et al. 2001, D'Antona et al.
  2002), and high stellar densities to favour mass-loss scenarios
  (Decressin et al.  2007; Huang \& Gies 2006). Such a comparison may
  therefore help to constrain the efficiency of EHB formation
  channels, provided that the present-day
  density and mass of the systems investigated have not experienced
  significant changes during the past, which could have been the case due to
  core collapse (Noyola \& Gebhardt 2006, de Marchi et al. 2007) or tidal stripping (e.g. Lee et al.  2007). To properly perform this comparison, deeper UV imaging data will be required that allow detection
  of UV intermediate-bright to faint GCs down to $M_V\simeq-$10 mag
  ($V\simeq 21.5$ mag at the Fornax distance).  In this respect, the
  outcomes of the HST observations in Cycle 15 (GO10901, PI O'Connell)
  of GCs belonging to NGC 1399 are highly anticipated.

\acknowledgements We are grateful to Caleb Scharf for providing us
with the source catalog of the Chandra Fornax survey. The work of
S.-C. R. was supported in part by KOSEF through the Astrophysical
Research Center for the Structure and Evolution of the Cosmos
(ARCSEC). { S.-J. Y. acknowledges support from the Basic Research
  Program (grant No. R01-2006-000-10716-0) and from the Korea Research
  Foundation Grant funded by the Korean Government (grant No.
  KRF-2006-331-C00134).}

\end{document}